\documentclass[conference]{IEEEtran}
\IEEEoverridecommandlockouts
\usepackage{cite}
\usepackage{amsmath,amssymb,amsfonts}
\usepackage{algorithmic}
\usepackage{graphicx}
\usepackage{subcaption}
\usepackage{textcomp}
\usepackage{xcolor}
\def\BibTeX{{\rm B\kern-.05em{\sc i\kern-.025em b}\kern-.08em
    T\kern-.1667em\lower.7ex\hbox{E}\kern-.125emX}}
\begin{document}

\title{A CARLA-based Simulation of Electrically Driven Forklifts 
\\

\thanks{\textbf{Funded by the Bavarian Ministry of Economic Affairs, Regional Development and Energy (Project
KAnIS, Grant No. DIK-1910-0016//DIK0103/01). }}
}

\author{\IEEEauthorblockN{1\textsuperscript{st} David Claus}
\IEEEauthorblockA{\textit{Fac. Engineering \& Computer Sciences} \\
\textit{Aschaffenburg Univ. of Applied Sciences}\\
Aschaffenburg, Germany \\
david.claus@boschrexroth.de}
\and
\IEEEauthorblockN{2\textsuperscript{nd} Christiane Thielemann}
\IEEEauthorblockA{\textit{Fac. Engineering \& Computer Sciences} \\
\textit{Aschaffenburg Univ. of Applied Sciences}\\
Aschaffenburg, Germany \\
christiane.thielemann@th-ab.de}
\and
\IEEEauthorblockN{3\textsuperscript{rd} Hans-Georg Stark}
\IEEEauthorblockA{\textit{Fac.  Engineering \& Computer Sciences} \\
\textit{Aschaffenburg Univ. of Applied Sciences}\\
Aschaffenburg, Germany \\
hans-georg.stark@th-ab.de}

}

\maketitle

\begin{abstract}

\textbf This paper presents the simulation of the operation of an electric forklift fleet within an intralogistics scenario. For this purpose, the open source simulation tool CARLA is used; according to our knowledge this is a novel approach in the context of logistics simulation. First, CARLA is used to generate and visualize a realistic 3D outdoor warehouse scenario, incorporating a number of randomly moving forklifts. In a next step, intralogistics transport tasks, such as pick-and-place, are simulated for the forklift fleet, including shortest-path finding. Furthermore, the capability to play back localization data, previously recorded from a "real" forklift fleet, is demonstrated. This play back is done in the original recreated environment, thereby enabling the visualization of the forklifts' movements. Finally, the energy consumption of the forklift trucks is simulated by integrating a physical battery model that generates the state of charge (SOC) of each truck as a function of load and activity.\newline
To demonstrate the wide range of possible applications for the CARLA simulation platform, we describe two use cases. The first deals with the problem of detecting regions with critically high traffic densities, the second with optimal placement of charging stations for the forklift trucks. Both use cases are calculated for an exemplary warehouse model.

\end{abstract}


\section{Introduction}
In recent years, the open-source simulation environment CARLA \cite{carla2} has become increasingly popular. CARLA is developed to support development, training and validation of autonomous driving cars. In addition to its primary function, CARLA offers the ability to integrate road maps and physical vehicle properties, which makes it superior to state-of-the-art simulation tools (ADD sources). In addition, the CARLA simulator employs the 3-D-creation tool Unreal Engine (Epic Games) for building realistic and customizable 3-D environments (buildings, etc.) for autonomous driving simulations. A remarkable feature of the Unreal Engine is its ability to produce high-quality 3-D graphics as well as providing physics and sensor simulation (cameras, LiDAR, radar) and simulating weather and lighting conditions. In addition, a Python interface facilitates the control of the CARLA simulator, the collection of data and the integration of new algorithms.\\In view of these remarkable features we propose here to adapt the powerful CARLA simulation platform for its use in industrial logistics tasks and, in particular, for simulating an autonomously driving forklift fleet in a realistic material handling scenario.
We state, that CARLA offers a wide range of possibilities, such as the support of the design of a warehouse layout, prediction and optimization of the energy consumption of the fleet, simulation of cooperation among the autonomously driving trucks, or simply visualization of forklift motion.  \\
Our contribution is organized as follows: In Sect. \ref{sec:SIM} we give an overview about our CARLA-platform architecture, forklift modeling, roadmap/geodata import and forklift motion simulation. Sect. \ref{sec:Battery} is devoted to the battery model mentioned above, the use cases are described in Sects. \ref{sec:CD} and \ref{sec:CS}. Summary and outlook are given in Sect. \ref{sec:Res}. 



\section{CARLA-based platform 
\label{sec:SIM}} 
\subsection{Platform architecture}\label{sec:CD2}

\begin{figure}[h]\centering
\includegraphics[width=9cm]{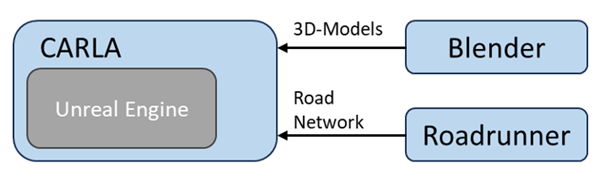}
        \caption{Platform Architecture/Toolchain}
        \label{fig:toolchain}
\end{figure}

The implemented simulation software is based on the toolchain depicted in Fig.\ref{fig:toolchain}.
\\
The Unreal Engine provides in particular the physics simulation environment, which in turn relies on the physics engine PhysX-software (Nvidia). 3D scenery and object surfaces are designed with Blender (Blender Foundation). The forklift shown in Fig. \ref{fig:SOC} is modeled using this tool. Roadrunner (Mathworks) is used to generate maps in an ASAM OpenDRIVE format, a standardized XML format to describe road networks, by setting up streets, adding crosswalks and defining the overall setup of the traffic rules.\\

Once the road network and 3D models have been imported into CARLA, the simulation is run using customised
Python scripts in a client-server architecture.

\subsection{Import of forklift model and GPS data for warehouse design \label{sec:CD4}}

\begin{figure}[h]

    \centering
        \includegraphics[width=\textwidth, height=5.2cm, keepaspectratio]{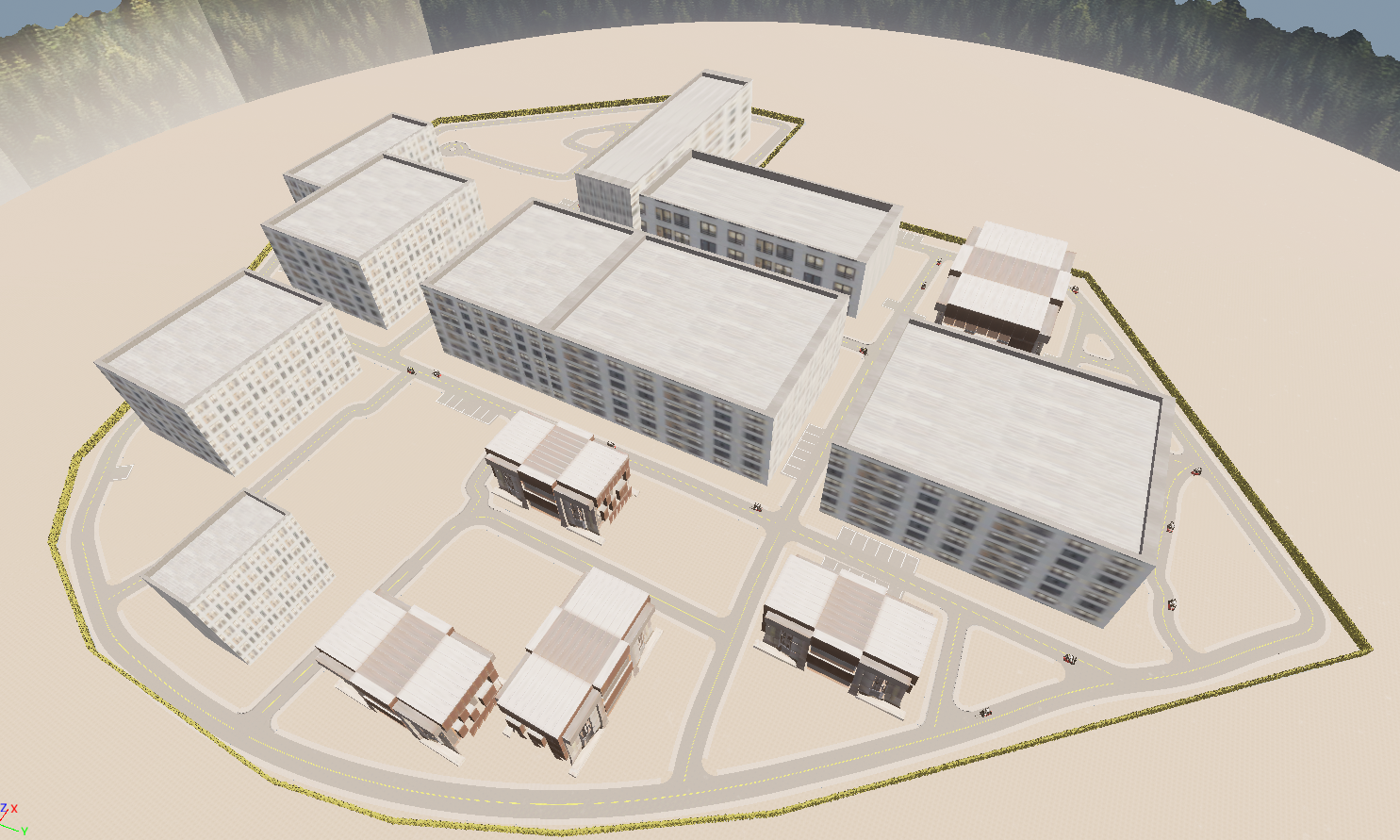}
        \caption{3D Visualization of a real hybrid warehouse in CARLA.}
        \label{fig:Virtual_Site}
   
\end{figure}

\begin{figure}[h]
    \centering
        \includegraphics[width=\textwidth, height=5.6cm,angle =90, keepaspectratio]{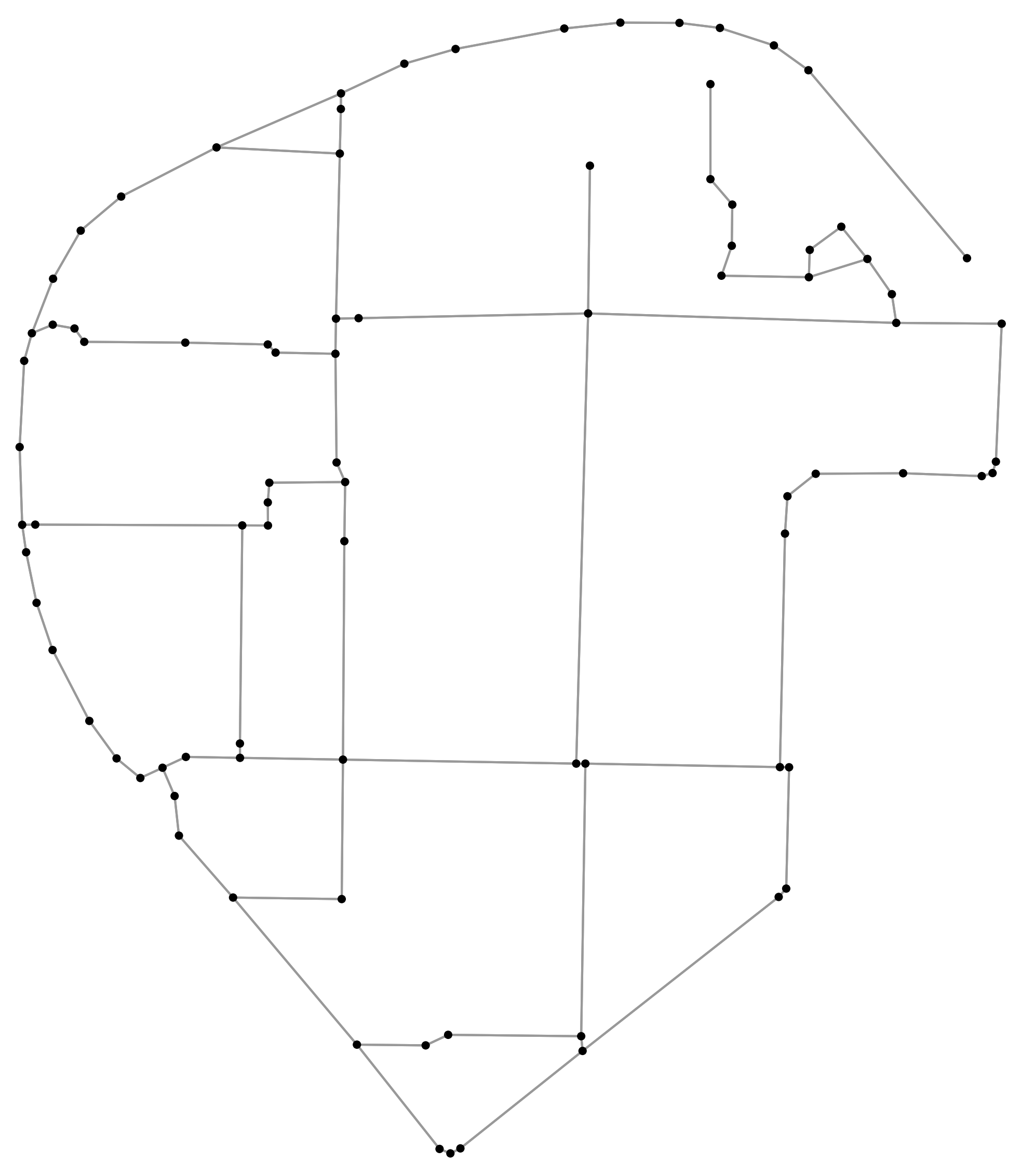}
        \caption{Graph  representing the road network of hybrid warehouse visualized in Fig. \ref{fig:Virtual_Site}, generated via Roadrunner.}
        \label{fig:roadnetwork}

\end{figure}

Fig. \ref{fig:Virtual_Site} gives an impression of the simulation frontend. \\ Fig. \ref{fig:roadnetwork} displays the corresponding graph consisting of vertices and edges. This road network plays a central role for planning and traversing trajectories of simulated forklifts on the roads. For smooth movements, like along curves, vertices need to be refined by a network of so-called waypoints. 

\subsection{Simulation\label{sec:CD3}}
To create the simulation of traffic during working hours, the real-world location data of several forklifts is used to identify areas that are regularly visited. These areas are then added to the simulation using custom-designed parking spots along the edges of the road network, see Fig. \ref{fig:SOC}. The parking spots represent starting position and destination of a variable number of simulated vehicles. Using the A* algorithm, the shortest route between start and destination is calculated. The destination of each simulated forklift varies and is either randomly chosen out of a list of unoccupied parking spots or predefined by the user. By predefining the destination, areas of high traffic load are created. In order to prevent the simulated forklifts from colliding with each other while driving from start to destination along the waypoints, a collision avoidance algorithm  using a top down approach with every vehicle knowing about the location and speed vector of all others, is used.\\

\section{Battery Model \label{sec:Battery}}
\begin{figure}[h!t]\centering
\includegraphics[width=9cm]{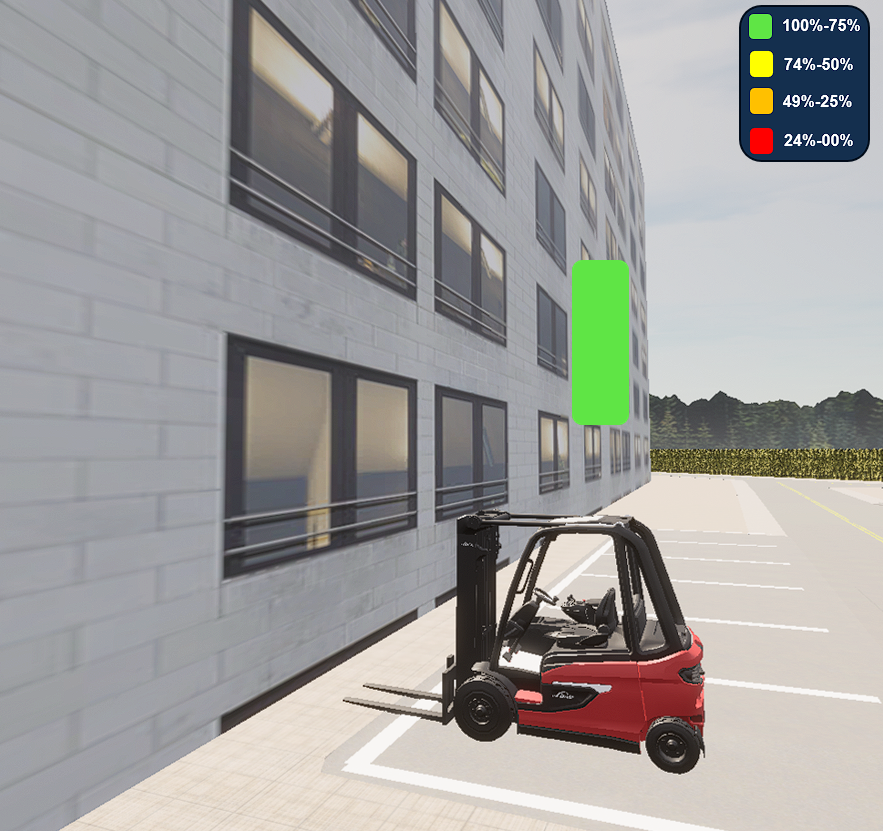}
\caption{Forklift in a parking position with visualization of its SOC (green bar corresponding to the value 100 \% - 75 \%). A SOC between 100 \% and 50 \% is considered to be a sufficient battery condition, while values below 50 \% suggest a trip to the charging station.  }
\label{fig:SOC}
\end{figure}

Our model is motivated by the force-balance methods described in \cite{Forces}. Given some trajectory $\Vec{x}(t) \; (t_0\le t\le t_e)$ of a forklifter, or a part thereof, the work $\Delta W$, which has to be drawn from the battery's energy along this trajectory, may be computed corresponding to
$$\Delta W=\int_{t_0}^{t_e} \Vec{F}(\Vec{x}(t)) \Vec{dx}$$
Here the total exerted force $\Vec{F}(\cdot)$ along the trajectory may be computed from driving forces, friction, steering resistance etc. and it can be split into horizontal components corresponding to driving/braking and vertical components corresponding to lifting/lowering loads and forks. All required forces can be deduced from the desired motion and from truck data like masses, friction coefficients etc. Empirical factors in the force equations \cite{Forces} may be determined from actual energy loss measurements performed on the real counterparts of the virtual trucks during, e.g., the VDI 2198 test cycle \cite{VDI2198} by replaying these measurements in the simulation environment and fitting the above parameters such that the simulated data correspond to the measured ones. 
Based on these considerations a runtime-information about the battery's SOC is computed and visualized as shown in Fig. \ref{fig:SOC}. 

\section{Applications}
\label{sec:app}
\subsection{Critical traffic density \label{sec:CD}}
\begin{figure}[h]\centering
\includegraphics[width=9cm]{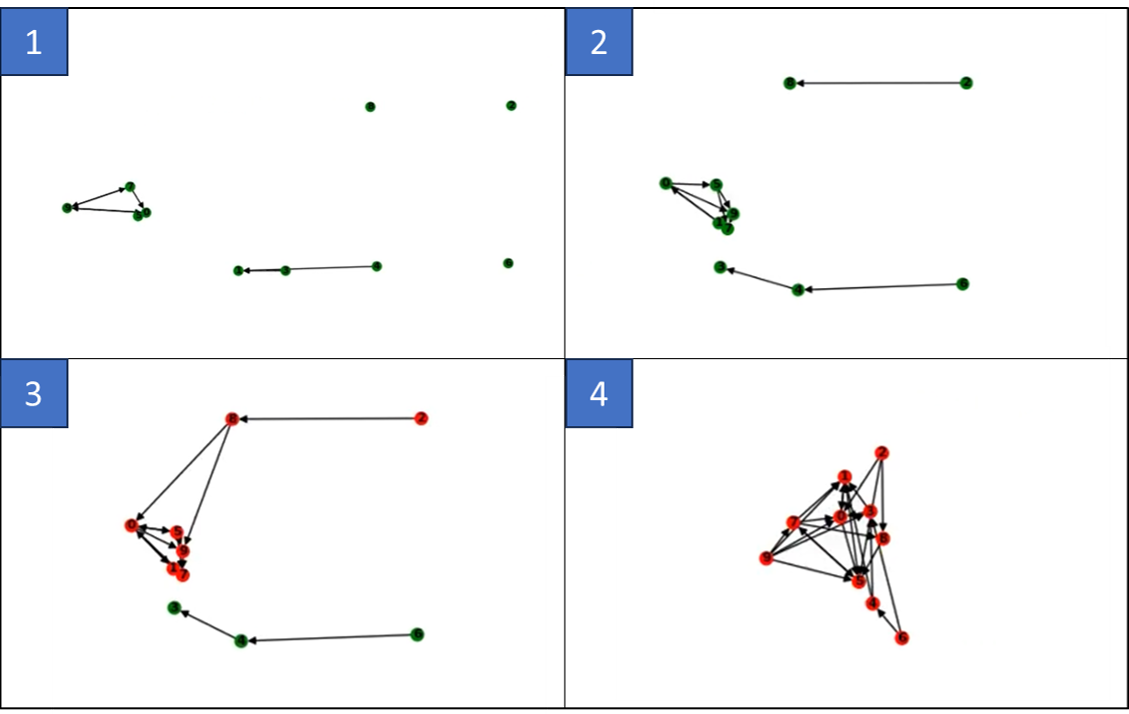}
\caption{Progression of a graph network over time with forklifts as dynamic vertices and the length of the shortest path between them as weighted edges. Two vertices form a cluster if the length of shortest path falls below a threshold. The nodes of a cluster are visualized in red if the average velocity of the cluster falls below a threshold. Note that the shortest path between two nodes is not the same for these node because of the layout of the map (right hand traffic). }
\label{fig:Graphs}
\end{figure}

During the planning phase of a new or the modification of an industrial plant, different aspects need to be considered. The work in this section is inspired by the approach of \cite{Wang} and aims to evaluate the traffic density during the planning phase of such a project and shall help to find critical areas or tasks that would otherwise lead to problems in the real-world implementation. 
We consider the industrial trucks/forklifts moving on the road network shown in the upper right part of Fig. \ref{fig:Virtual_Site} as vertices of a (dynamical) graph, for a short introduction to graphs we refer to, e.g., \cite{ortega, Shuman1}. Those vertices are linked by edges, which are weighted by the shortest distance between them, calculated on the roadmap graph. This is a standard tool in graph analytics, it is based on Dijkstra's algorithm \cite{SD}. Fig. \ref{fig:Graphs} depicts the progression of the graph over time (from 1 to 4). The reader should not be confused by the fact, that the underlying road network is for reasons of clarity not shown, nevertheless vertices are always positioned on that road network.

The above graph is split into separate regional clusters according to the following rule: A vertex $v_k$ belongs to a cluster $\mathcal{C}$ if there is at least one cluster member $v_m\in \mathcal{C}$ such that $d(v_k,v_m)\le t$, where $d(v_k,v_m)$ denotes the above-mentioned shortest distance between both vertices and $t$ is some threshold. As the motion evolves, Fig. \ref{fig:Graphs} shows such clusters. Note that there are directed edges in the clusters; the reason is, that due to traffic rules and/or one-way-sections on the road network the shortest distance $d(v_k,v_m)$ from $v_k$ to $v_m$ might be different from the shortest distance $d(v_m,v_k)$ from $v_m$ to $v_k$. 

At every time step the average velocity inside the individual clusters may be calculated and if this average falls below some predefined threshold the cluster is highlighted as critical. With this information, forklift task distributions or plant areas like narrow corners that lead to a critical traffic density can be detected.

\subsection{Placement of charging stations \label{sec:CS}}
\begin{figure}[h]\centering
\includegraphics[width=4.4cm]{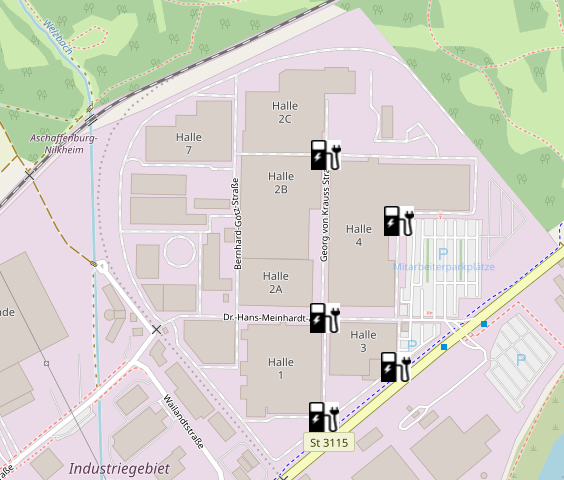}
\includegraphics[width=4.3cm]{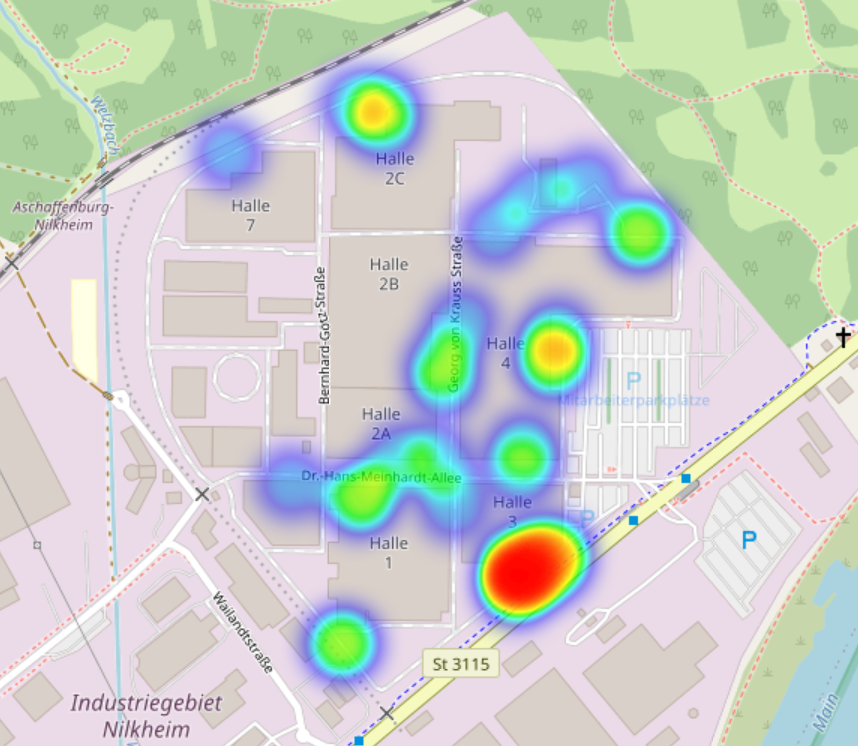}
\caption{Optimized positions of charging stations ($k=5$) (left). Right: Result of a traffic heatmap calculated from recorded  forklift motion data. Quoted from \cite{PageRank}}%
\label{fig:opl_result}%
\end{figure}

The optimal placement of charging stations for electrically driven vehicles is a well-known and important problem in order to provide the fleet with smooth and reliable power supply, see, e.g., \cite{EVCSP2}.
In \cite{PageRank} an algorithm has been described, which computes optimal positions for charging stations based on the analysis of previously recorded forklift motion data. Thus the algorithm depends on a set of $N$ forklift trajectories $\{\Vec{x}_n(t) \; | \; t_0\le t\le t_e, n=1,\ldots,N\}$, where $[t_0,t_e]$ denotes the time interval considered and $n$ numbers the forklifts. A further input parameter to the algorithm is the maximum number $k$ of possible charging station positions to be computed by the algorithm. In Fig. \ref{fig:opl_result} a result is shown, where real-life forklift motion data, recorded in the exemplary warehouse for a relatively small number of forklifts over a certain period of time, have been fed into the algorithm. The computed positions of charging stations are compared with a heat map analysis \cite{heatmapanalyse} of the recorded traffic data; obviously there is a reasonable correlation of the computed positions to heatmap maxima.\\

\begin{figure}[h]\centering
\includegraphics[width=3.5cm]{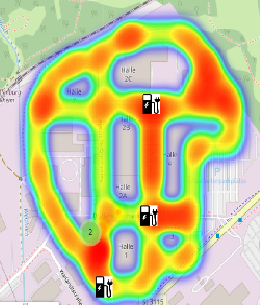}
\caption{Optimized positions of charging stations superimposed with traffic heatmap. The traffic heatmap is calculated from simulated forklift motion data.}
\label{fig:Simopl_result}%
\end{figure}

With the CARLA-Simulation environment, described in this paper, available, we are now able to perform the same type of analysis with a scalable number of forklifts. We rebuilt the warehouse site and generated simulated motion data of a forklift fleet as described above. Then, via the motion data import/export interface of our simulation environment, we fed the motion data into the placement algorithm. The results are again compared to a corresponding heatmap and shown in Fig. \ref{fig:Simopl_result}.

It is rather obvious that the procedure described in this section is useful for, e.g., planning tasks such as simulating and optimizing intralogistic tasks scheduled for future periods. Considering our battery model from above, it should be possible to enhance the algorithm such that the individual SOC's of the forklifts (c.f. Fig. \ref{fig:SOC}) also are taken into account. Moreover, the optimized supply of electrical power is actually a special version of the general problem of matching sources and sinks of ressources. Thus an adaptation of the algorithm to, e.g., pick-and-drop-tasks might also be feasible.

\section{Result, Discussion and Future Work \label{sec:Res}}

In this paper, we present a simulation platform based on CARLA that enables the modeling, visualization and simulation of an electric forklift fleet in a customizable 3D intralogistics scenario.  \\
As an example, we model a  customizable 3D warehouse that includes a scalable number of realistic forklifts. 
On the one hand, the ability to replay "real" recorded motion data in the warehouse model is demonstrated. On the other hand, the behavior of a forklift fleet is simulated and typical intralogistics tasks, such as pick-and-place, are performed, enabling the generation of \textit{in silico} data. The latter is valuable for further analysis tasks, as demonstrated in section \ref{sec:app}, where critical traffic densities and  placement of charging stations are calculated based on simulated data. \\
Continuous monitoring of the battery's SOC (see section \ref{sec:Battery}) allows for extending the shortest distance-based algorithms with criteria related to the energy resources of individual vehicles.



In conclusion, the potential of the intralogistics-simulation environment is impressive. Its scalability allows for the planning of logistics tasks scheduled for future periods via simulation runs. The detection of critical densities and the computation of resources, such as charging stations, illustrates the usability of the tool in this context, as well as its use as a support tool for performance analysis when new warehouse sites are configured.

\section{Acknowledgement}
The authors are grateful to Linde Material Handling GmbH, the industrial partner in this project, for providing truck data and essential advice. The first named author is also thankful for the financial support provided by Bosch Rexroth AG.

\bibliographystyle{IEEEtran}
\bibliography{references}

\end{document}